\documentstyle[11pt,twoside,newpasp,epsf]{article}
\def\edcomment#1{\iffalse\marginpar{\raggedright\sl#1\/}\else\relax\fi}
\reversemarginpar

\newcommand{\sak}{\mbox{V\,4334\,Sgr}}

\newcommand{\jahre}{\, \mathrm{yr}}

\begin{document}
\title{Past and future evolution  of Sakurai's object}

 \author{Falk Herwig}
\affil{University of Victoria, Victoria, BC, Canada;
  fherwig@uvastro.phys.uvic.ca}

\begin{abstract}
We present a stellar evolution model sequence of the past and future evolution
of the post-AGB born again star Sakurai's object (\sak). In
order to match the short evolutionary time scale we have to assume
that the convective ingestion of hydrogen-rich envelope material into the
He-flash convection zone proceeds slower than predicted by the
mixing length theory. For the future we predict a swift
second evolution through the central star region before a second
born-again evolution occurs.  
\end{abstract}

The fast evolving born again asymptotic giant branch
(AGB) star \sak\ provides a unique opportunity to test stellar
evolution and nucleosynthesis models of the very late thermal pulse
(VLTP) which occurs in about 10$\%$ of all central stars of planetary
 nebulae (CSPN) (Iben \& McDonald, 1995)\nocite{iben:95b}.  \sak\
 displayed an extraordinary fast evolution from the
 pre-white dwarf (WD) stage to the AGB in only three years
 \nocite{duerbeck:00} (Duerbeck et al., 2000). In addition a  peculiar
 real-time abundance evolution has been documented for this star by Asplund et
 al.\ (1999).\nocite{asplund:99a}  We address both of these properties
 by exploring theoretical models. For example, Herwig \& Langer (2001)
 \nocite{herwig:00f} proposed a new
nucleosynthesis mechanism for the lithium overabundance of \sak,
because the well known hot-bottom burning mechanism which operates in
massive AGB stars had to be excluded here. 
However, the latest VLTP stellar evolution model by Herwig et al.\ (1999)
\nocite{herwig:99c}  could not match the observed fast evolution of
\sak\ although considerable effort had been put into a consistent
numerical method for simultaneous mixing and
nucleosynthesis. Herwig (2001) \nocite{herwig:01a} assumed 
that convective mixing of envelope material (cold fuel)
into the He-flash convection zone is retarded with respect to the
mixing-length theory (MLT) velocity. In that case the p-capture nuclear energy is
released closer to the surface which leads to a faster
inflation of the outermost layers.  With this model the observed born
again evolution time scale of V 4334 Sgr can be matched (Fig.\,1). 
It should be noted, that MLT was
never designed to apply to the ingestion of cold fuel into a
convectively unstable nuclear burning zone.

The new model  reproduces the C-isotopic and the C/N elemental ratio of
   Sakurai's object.  Much more detailed information is
   available from the spectroscopic study of Asplund et al.\ (1999),
   and the observed iron-deficiency has already been addressed by Herwig,
   Lugaro \& Werner (this conference).  
An important feature of the new VLTP model sequence is the superposition of two born again
phases (Fig.\,1). The first, fast return to the AGB is
driven by the H-flash fueled by protons ingested into deeper layers from the
envelope. After the star returns to the CSPN stage it is predicted to
follow a second, He-flash  driven born again
evolution which proceeds on the more familiar time scale of the order
of $\sim 100\jahre$. The current location of
   Sakurai's object is at the end of the 1$^{\rm st}$ return to the
   AGB. The predicted second evolution through the 
   CSPN phase should be noticeable on a timescale of decades and we
   therefore expect that there is more to learn from Sakurai's 
   object in our lifetime.
\begin{figure}
\label{fig1}
\plotone{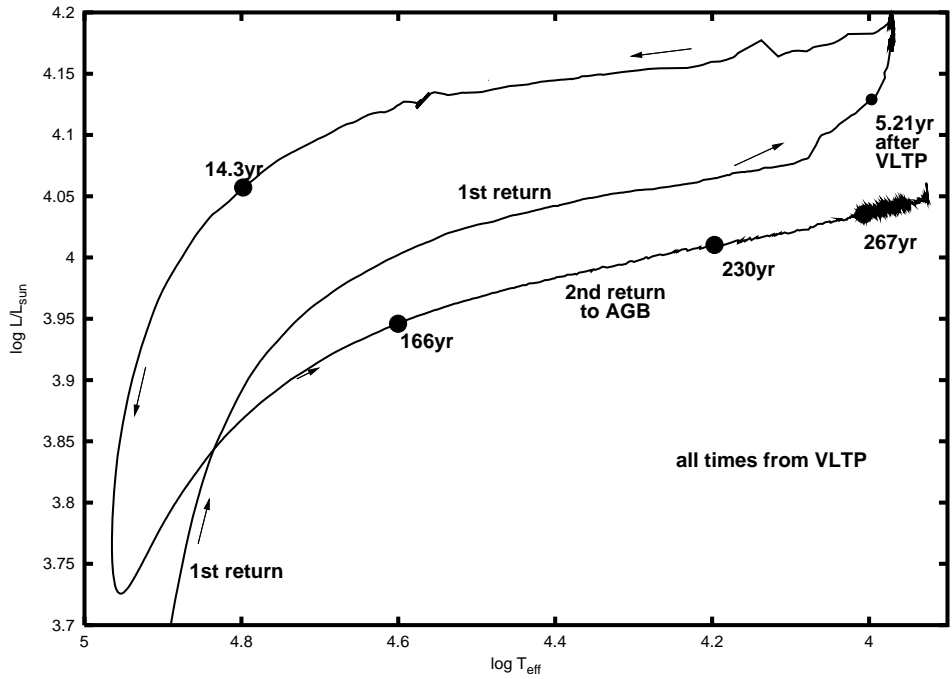}  
\caption{Post-AGB stellar evolution model sequence following a  very late thermal
  pulse for comparison with Sakurai's object.}
\end{figure}

\noindent
\textbf{Acknowledgments:} I would like to thank
D.A.\ VandenBerg for support through his Operating Grant from the Natural
Science and Engineering Research Council of Canada.

\end{document}